\documentstyle[aps,eqsecnum]{revtex}
\def\G{\Gamma}

\def\w{\Omega}
\def\C{{\bf C}}
\def\c{{\cal C}}
\def\S{{\cal S}}

\def\la{\langle}
\def\ra{\rangle}
\def\be{\nopagebreak[3]\begin{equation}}
        \def\ee{\end{equation}}
        \def\ba{\nopagebreak[3]\begin{eqnarray}}
        \def\ea{\end{eqnarray}}

\def\d{{\rm d}}

\newcommand{\teta}{\rlap{\lower2ex\hbox{$\,\tilde{}$}}\eta{}}

\begin{document}

\preprint{\vbox{\baselineskip=12pt \rightline{ICN-UNAM-02/10}
\rightline{hep-th/0207088} }}

\draft
\title{On the Schr\"odinger Representation for a \\
Scalar Field on Curved Spacetime}
\author{Alejandro Corichi${}^{1,2,3}$,
Jer\'onimo Cortez${}^1$ and Hernando Quevedo${}^1$\thanks{corichi,
cortez, quevedo@nuclecu.unam.mx}}
\address{1. Instituto de Ciencias Nucleares\\
Universidad Nacional Aut\'onoma de M\'exico\\
A. Postal 70-543, M\'exico D.F. 04510, M\'exico \\
}
\address{2. Department of Physics and Astronomy\\
University of Mississippi, University, MS 38677, USA\\
}
\address{3. Perimeter Institute for Theoretical Physics\\
35 King Rd. North, Waterloo, Ontario  N2J 2W9, Canada\\
}

 \maketitle

\begin{abstract}
It is generally known that linear (free) field theories are one of
the few QFT that are exactly soluble. In the Schr\"odinger
functional description of a scalar field on flat Minkowski
spacetime and for flat embeddings, it is known that the usual Fock
representation is described by a Gaussian measure. In this paper,
arbitrary globally hyperbolic space-times and embeddings of the
Cauchy surface are considered. The classical structures relevant
for quantization are used for constructing the Schr\"odinger
representation in the general case. It is shown that in this case,
the measure is also Gaussian. Possible implications for the
program of canonical quantization of midisuperspace models are
pointed out.
\end{abstract}
\pacs{03.70.+k, 04.62.+v}

\section{Introduction}
\label{sec:1}

The quantum theory of a free real scalar field is probably the
simplest field theory system. Indeed, it is studied in the first
chapters on most field theory textbooks \cite{textbook}. The
language used for these treatments normally involves Fourier
decomposition of the field and creation and annihilation operators
associated with an infinite chain of harmonic oscillators.
Canonical quantization is normally performed by representing these
operators on Fock space  and implementing the Hamiltonian
operator. However, from the perspective of ``canonical
quantization", where one starts from a classical Poisson algebra
and performs a quantization of the system, this procedure is not
always transparent. These issues have been addressed by Wald who
motivated by the study of quantum fields on curved spacetimes
deals with the process of quantization, starting from a classical
algebra of observables and constructing representations of them on
Hilbert spaces \cite{wald2}. Furthermore, Wald develops the
quantum theory of a scalar field, and extends the formalism to an
arbitrary globally hyperbolic curved manifold. His construction
is, however, restricted to finding a representation on Fock space,
or as is normally known, the {\it Fock representation}.

On the other hand, the usual presentation of elementary quantum
mechanics pays a lot of attention to the Schr\"odinger
representation, where quantum states are represented by functions
on configuration space. Thus, the construction of the functional
Schr\"odinger representation for fields seems to be a natural step
in this direction. It is therefore unsettling that a complete and detailed 
treatment for curved spacetimes does not seem to be available in
the literature. The purpose of this paper is to fill this gap. The
Schr\"odinger representation for fields on {\it Minkowski}
spacetime, where an inertial slicing of the spacetime is normally
introduced, has been previously studied (for reviews see
\cite{reviews}). This functional viewpoint, even when popular in
the past, is not widely used, in particular since it is not the
most convenient one for performing calculations of physical
scattering processes in ordinary QFT.\footnote{However, it has
been successfully used for proving a variety of results that do
not need dynamical information \cite{jackiw}.}

However, from the conceptual viewpoint, the study of the
Schr\"odinger representation in field theory is extremely
important and has not been, from our viewpoint widely acknowledged
(however, see \cite{jackiw}). This is specially true since some
symmetry reduced gravitational system can be rewritten as the
theory of a scalar field on a {\it fiducial, flat}, background
manifold. In particular, of recent interest are the polarized
Einstein-Rosen waves \cite{einstein} and Gowdy cosmologies
\cite{gowdy,ccq}. The Schr\"odinger picture is then, in a sense,
the most natural representation from the viewpoint of canonical
quantum gravity, where one starts from the outset with a
decomposition of spacetime into a spatial manifold $\Sigma$
``evolving in time". Therefore, it is extremely important to have
a good understanding of the mathematical constructs behind this
representation and its relation to the Fock representation.

In this regard, there seems to be an apparent tension in the
construction of the Schr\"odinger representation for a scalar
field. On the one hand, if one follows a systematic approach to
quantization, as outlined for instance in \cite{tate,marolf} one
can, without difficulties arrive at the ``ordinary" representation
of the elementary quantum operators \cite{reviews,jackiw,dvlong}, where
the quantum measure is ``homogeneous". However, we know from the
more rigorous treatments of the subject \cite{glimm,baez,velh}, that a
consistent quantization should involve a non-homogenous, Gaussian
measure, and therefore a non-standard representation of the
(momentum) operators. This seems to indicate that one needs
additional ``dynamical" input within the algebraic quantization
procedure \cite{tate,marolf}.

The purpose of this paper is  to systematically construct the
functional quantum theory and extend the rigorous formalism of
\cite{glimm,baez} to arbitrary embeddings of the Cauchy surface
and to arbitrary curved spacetimes, in the spirit of \cite{wald2}.
The emphasis we shall put regarding the relevant structures will
allow us to achieve this goal. The generalization that we will
construct in this paper will be at two levels. 
To be specific, we firstly deal
with the existing ambiguity in the quantization of a scalar field,
already recognized in the Fock quantization \cite{wald2}.
Furthermore,  the infinite freedom in the choice of embedding of
the Cauchy surface is considered. We  find that the measure is
always Gaussian in an appropriate sense and that it can be written
in a simple way. Even when straightforward, these results have
not, to the best of our knowledge, appeared elsewhere. As an
offspring, they provide the required language for the systematic
treatment of symmetry reduced 
models within canonical quantum
gravity \cite{einstein,gowdy,ccq}, and provide an elegant solution
to the apparent tension mentioned above.

The structure of the paper is as follows. In Sec.~\ref{sec:2} we
recall basic notions from the classical formulation of a scalar
field.  A discussion of the Schr\"odinger representation and
construction of the functional description, unitary equivalent to
a given Fock representation, is the subject of
Section~\ref{sec:3}. This is the main section of the paper.  We
end with a discussion in Sec.~\ref{sec:4}.

In order to make this work accessible not only to specialized
researchers in theoretical physics, we have intentionally avoided
going into details regarding functional analytic issues and other
mathematically sophisticated constructions. Instead, we refer to
the specialized literature and use those results in a less
sophisticated way, emphasizing at each step their physical
significance. This allows us to present our results in a
self-contained fashion.

\section{Classical Preliminaries}
\label{sec:2}

In this section we recall the classical theory of a real, linear
Klein-Gordon field $\phi$ with mass $m$ propagating on a
4-dimensional, globally hyperbolic spacetime
$(^{\mbox{\tiny{4}}\!}M,g_{ab})$. As is well-known, global
hyperbolicity implies that $^{\mbox{\tiny{4}}\!}M$ has topology
${\bf{R}}\times \Sigma$,  and can be foliated by a one-parameter
family of smooth Cauchy surfaces diffeomorphic to $\Sigma$. Hence,
we can  perform a 3+1 decomposition of the spacetime of the form
${\bf{R}}\times \Sigma$ and consider arbitrary embeddings of the
surface $\Sigma$ into $^{\mbox{\tiny{4}}\!}M$.

This section has two parts. In the first one, we recall the
canonical treatment of the scalar field, together with the
observables that are relevant for quantization. In the second
part, we introduce a classical construct that is needed for
quantization, namely a complex structure on phase space.

\subsection{Canonical phase space and observables}
\label{sec:2.a}

The phase space of the theory can be alternatively described by
the space $\G$ of Cauchy data (in the canonical approach), that
is, $\{(\varphi , \pi)\vert \, \varphi : \Sigma \to {\bf{R}},\,
\pi: \Sigma \to {\bf{R}}; \, \varphi ,
 \pi \in C^{\infty}_{0}(\Sigma)\}${\footnote{The class of functions
 comprised by Schwartz space is most commonly chosen for quantum
 field theory in Minkowski spacetime. However, the notion of Schwartz
  space is not extendible in any obvious way to more general manifolds
  \cite{wald2,draft}. Hence, we shall define $\G$ to consist of initial
  data which are smooth and of compact support on $\Sigma$.}}, or by
  the space $V$ of smooth solutions to the Klein-Gordon equation which
  arises from initial data on $\G$ (in the covariant formalism) \cite{wald2}.
  Note that, for each embedding $T_{t}:\Sigma \to {^{\mbox{\tiny{4}}\!}M}$,
  there exists an isomorphism ${\cal{I}}_{t}$ between $\G$ and $V$. The key
  observation is that there is a one to one correspondence between a pair
  of initial data of compact support on $\Sigma$, and solutions to the
  Klein-Gordon equation on $^{\mbox{\tiny{4}}\!}M$. That is to say:

Given an embedding $T_{t_{0}}$ of $\Sigma$ as a Cauchy surface
$T_{t_{0}}(\Sigma)$ in $^{\mbox{\tiny{4}}\!}M$, the (natural)
isomorphism ${\cal{I}}_{t_{0}}:\G\to V$ is obtained by taking a
point in $\G$ and evolving from the Cauchy surface $T_{t_{0}}(\Sigma)$
 to get a solution of $(g^{ab}\nabla_{a}\nabla_{b}-m^{2})\phi =0$.
 That is, the specification of a point in $\G$ is the appropriate initial
 data for determining a solution to the equation of motion. The inverse
map, ${\cal{I}}_{t_{0}}^{-1}:V \to \G$, takes a point $\phi \in V$
and finds the Cauchy data induced on $\Sigma$ by virtue of the
embedding $T_{t_{0}}$: $\varphi=T_{t_{0}}^{*}\phi$ and
$\pi=T_{t_{0}}^{*}(\sqrt{h}\pounds_{n}\phi)$, where $\pounds_{n}$
is the Lie derivative along the normal to the Cauchy surface
$T_{t_{0}}(\Sigma)$ and $h$ is the determinant of the induced
metric on such a surface. Note that the phase space $\G$ is of the
form $T^{*}{\c}$, where the classical configuration space ${\c}$
is comprised by the set of smooth real functions of compact
support on $\Sigma$.

Since the phase space $\G$ is a linear space, there is a
particular simple choice for the set of fundamental observables.
We can take a global chart on $\G$ and we can choose the set of
fundamental observables to be the vector space generated by {\it
linear} functions on $\Gamma$. More precisely, classical
observables for the space $\G$ can be constructed directly by
giving smearing functions on $\Sigma$. We can define linear
functions on $\G$ as follows: given a vector $Y^\alpha$ in $\G$ of
the form $Y^\alpha=(\varphi,\pi)^\alpha$ (note that due to the
linear nature of $\G$, $Y^{\alpha}$ also represents a vector in $T \G$)
and a pair $\lambda_\alpha=(-f,-g)_\alpha$, where $f$ is a scalar
density and $g$ a scalar, we define the action of $\lambda$ on $Y$
as, \be \label{observ} F_\lambda(Y)=- \lambda_\alpha
Y^\alpha:=\int_\Sigma(f\varphi+g\pi)\,\d^3x \, .\ee Now, since in the
phase space $\G$ the symplectic structure $\Omega$ takes the
following form, when acting on vectors $(\varphi_1,\pi_1)$
and$(\varphi_2,\pi_2)$, \be
\Omega([\varphi_1,\pi_1],[\varphi_2,\pi_2])=\int_\Sigma(\pi_1\varphi_2-
\pi_2\varphi_1)\,\d^3x \, ,\label{symp2} \ee then we can write the
linear function (\ref{observ}) in the form
$F_\lambda(Y)=\Omega_{\alpha\beta}\lambda^\alpha
Y^\beta=\Omega(\lambda,Y)$, if we identify
$\lambda^\beta=\Omega^{\beta\alpha}\lambda_\alpha=(-g,f)^\beta$.
That is, the smearing functions $f$ and $g$ that appear in the
definition of the observables $F$ and are therefore naturally
viewed as a 1-form on phase space, can also be seen as the vector
$(-g,f)^\beta$. Note that the role of the smearing functions is
interchanged in the passing from a 1-form to a vector. Of
particular importance for what follows is to consider {\it
configuration} and {\it momentum} observables. They are particular
cases of the observables $F$ depending of specific choices for the
label $\lambda$. Let us consider the ``label vector''
 $\lambda^\alpha=(0,f)^\alpha$, which would be normally regarded
as a vector in the ``momentum'' direction. However, when we
consider the linear observable that this vector generates, we get,
\begin{equation}
\varphi[f]:=\int_\Sigma \d^3\! x\,f\,\varphi\, .\label{observ2}
\end{equation}
Similarly, given the vector  $(-g,0)^\alpha$ we can construct,
\begin{equation}
\pi[g]:=\int_\Sigma\d^3\! x\,g\,\pi .\label{observ3}
\end{equation}

Note that any pair of test fields $(-g,f)^\alpha\in \G$ defines a
linear observable, but they are `mixed'. More precisely, a scalar
$g$ in $\Sigma$, that is, a pair $(-g,0)\in \G$ gives rise to a
{\it momentum} observable $\pi[g]$ and, conversely, a scalar
density $f$, which gives rise to a vector $(0,f) \in \G$ defines a
{\it configuration} observable $\varphi[f]$. In order to avoid
possible confusions, we shall make the distinction between {\it
label vectors} $(-g,f)^\alpha$ and {\it coordinate vectors}
$(\varphi,\pi)^\alpha$.

It is important to emphasize that the symplectic structure
provides the space of classical observables with an algebraic
structure via the Poisson bracket (PB). If we are now given
another label $\nu$, such that $G_\nu(Y)=\nu_\alpha Y^\alpha$, we
can compute the Poisson Bracket \be
\{F_\lambda,G_\nu\}=\Omega^{\alpha\beta}\nabla_\alpha
F_\lambda(Y)\nabla_\beta G_\nu(Y)=
\Omega^{\alpha\beta}\lambda_\alpha\nu_\beta \, . \ee Since the two-form
is non-degenerate we can rewrite it as
$\{F_\lambda,G_\nu\}=-\Omega_{\alpha\beta}\lambda^\alpha\nu^\beta$.
Thus, \be \label{poissonbra} \{\Omega(\lambda,Y),\Omega(\nu,Y)\}=-
\Omega(\lambda,\nu) \, . \ee

\subsection{Complex structure}
\label{sec:2.b}

Now, in order to provide the canonical approach with the
mathematical structure that encodes the inherent ambiguity in QFT,
and exactly in the same sense as it is done for the Fock
quantization, we have to introduce at the classical
level  a complex structure on the canonical phase space $\G$,
compatible with the symplectic structure (\ref{symp2}). Recall
that in the Fock picture which is naturally constructed from the
covariant phase space $V$, the complex structure is compatible with
the symplectic form defined on that space (a complex structure is
a linear mapping such that $J^2=-1$). Now, for the canonical phase
space, given an embedding $T_{t_{0}}:\Sigma \to
{^{\mbox{\tiny{4}}\!}M}$ and a complex structure on $V$, there is
one complex structure on $\G$ induced by virtue of the symplectic
map ${\cal{I}}_{t_{0}}$. On the symplectic vector space $(\G, \w)$
with coordinates $(\varphi,\pi)$, the most general form of the
complex structure $J$ is given by
\begin{equation}
\label{est.com}
-J_{\Gamma}(\varphi , \pi)=(A \varphi + B \pi ,C\pi +D\varphi) \, ,
\end{equation}
where $A,B,C$ and $D$ are linear operators  satisfying the following
relations \cite{6}:
\begin{equation}
\label{relaciones1} A^{2}+BD=-{\bf{1}} \:\: , \:\:\:
C^{2}+DB=-{\bf{1}} \:\: , \:\:\: AB+BC=0 \:\: , \:\:\: DA+CD=0 \, .
\end{equation}
The inner product $\mu_{\Gamma}(\, \cdot \, , \, \cdot \,)=\w(\,
\cdot \, ,-J_{\G}\, \cdot \,)$ in terms of these operators is
explicitly given by
\begin{equation}
\label{pint-operadores} \mu_{\Gamma}((\varphi_{1} ,
\pi_{1}),(\varphi_{2}, \pi_{2}))= \int_{\Sigma}\d^3x \,(\pi_{1} B
\pi_{2} + \pi_{1} A \varphi_{2} - \varphi_{1} D
\varphi_{2}-\varphi_{1} C \pi_{2} ) \, ,
\end{equation}
for all pairs $(\varphi_{1} , \pi_{1})$ and $(\varphi_{2},
\pi_{2})$. As $\mu_{\Gamma}$ is symmetric, then the linear
operators  should also satisfy \cite{6}
\begin{equation}
\label{relaciones2} \int_{\Sigma} fBf' = \int_{\Sigma} f'Bf \:\: ,
\:\:\: \int_{\Sigma} gDg' = \int_{\Sigma} g'D g  \:\: , \:\:\:
\int_{\Sigma}fAg = -
\int_{\Sigma} gCf \, ,
\end{equation}
where $g,g' \in C^{\infty}_{0}(\Sigma)$ are scalars, and $f,f' \in
C^{\infty}_{0}(\Sigma)$ are scalar densities of weight one.

As mentioned before, given an embedding $T_{t_{0}}$ of $\Sigma$
there is a one to one correspondence between complex structures on
$V$ and $\G$. That is, if we have a particular isomorphism
${\cal{I}}_{t_{0}}$, the complex structure induced on $\G$ by $J_{V}$,
a particular complex structure on the covariant phase space, is given
by $J_{\G}={\cal{I}}_{t_{0}}^{-1}J_{V}{\cal{I}}_{t_{0}}$. This relation
 and the general form (\ref{est.com}) implies that
\be \label{part-realiza} A\varphi + B\pi =
-T^{*}_{t_{0}}[J_{V}\phi]\: , \:\:\: C\pi + D\varphi =
-T^{*}_{t_{0}}[\sqrt{h}\pounds_{n}(J_{V}\phi)] \, , \ee where
$\phi={\cal{I}}_{t_{0}}(\varphi,\pi)$ (i.e., $\phi$ is the
solution to the Klein-Gordon equation which arises from the Cauchy
data pair $(\varphi,\pi)$) Thus, the particular realization of the
operators $A,B,C$ and $D$ will be different for different
embeddings $T_{t}$ of $\Sigma$.

\section{Schr\"odinger Representation}
\label{sec:3}

In this section, we turn our attention to the Schr\"odinger
representation. In contrast to the Fock case, which is most
naturally stated and constructed in a covariant framework
\cite{wald2}, this construct relies heavily on a Cauchy surface
$\Sigma$, since its most naive interpretation is in terms of a
``wave functional at time $t$".

Let us begin by looking at the classical observables that are to be quantized,
and in terms of which the CCR are expressed. Since the vector space of
elementary classical variables is ${\S}={\rm{Span}}\{1,\varphi[f],\pi[g]\}$
and  there is an abstract operator $\hat{F}$ in the free associative
algebra generated by $\S$ associated with each element $F$ in $\S$,
then we have that the basic quantum operators are $\hat{\varphi}[f]$
and $\hat{\pi}[g]$. The canonical commutation relations arises by
imposing the Dirac quantization condition on the basic quantum operators,
thus from (\ref{observ2}),  (\ref{observ3}) and (\ref{poissonbra}), the
CCR  read $[\hat{\varphi}[f],\hat{\pi}[g]] =i\hbar\int \d^3\!x\; fg \,
{\rm \hat{I}}$ (For a general discussion and details about the set ${\S}$
 and the steps for passage to quantum theory, in the canonical framework,
  see \cite{tate}).

Now, the Schr\"odinger representation, at least in an intuitive level, is
to consider `wave functions' as function(al)s of
$\varphi$. More precisely, the Schr\"odinger picture consists in representing 
the abstract operators $\hat{\varphi}[f]$ and $\hat{\pi}[g]$ as operators
in ${\cal H}_{\rm s}:=L^2(\overline{\cal C},\d\mu)$, where a state would
be represented by a function(al) $\Psi[\varphi]: \overline{\cal C}
\rightarrow \C$, with the appropriate ``reality conditions", which
in our case means that these operators should be Hermitian.

As a first trial, inspired and tempted by the names
``configuration'' and ``momentum'', one can try to represent the
corresponding operators as is done in ordinary quantum mechanics,
namely, by multiplication and derivation, respectively. However,
one must be careful since, in contrast to ordinary quantum
mechanics, the configuration space of the theory is now infinite
dimensional and Lebesgue type measures are no longer available
(The theory of measures on infinite dimensional vector spaces has
some subtleties, among which is the fact that well defined
measures should be {\it{probability}} measures \cite{draft,yama}.
A uniform measure would not have such a property). As a consequence
of the intimate relation between measure and operator
representation, and in order to reflect the nonexistence of a
homogeneous measure in a consistent way, we have to modify a bit
the simplest extension (suggested by ordinary quantum mechanics)
and represent the basic operators, when acting on functionals
$\Psi[\varphi]$, as follows \be
(\hat{\varphi}[f]\cdot\Psi)[\varphi]:=\varphi[f]\,\Psi[\varphi]\,
, \label{comfop} \ee and \be
(\hat{\pi}[g]\cdot\Psi)[\varphi]:=-i\hbar\int_{\Sigma}
\d^3\!x\;g(x)
 {{\delta \Psi}\over{\delta \varphi(x)}} + {\rm
multiplicative\; term}\, .\label{momop} \ee where the second term
in (\ref{momop}), depending only on configuration variable, is
precisely there to render the operator self-adjoint when the
measure is different from the ``homogeneous" measure, and depends
on the details of the measure. The first thing to note is that the
representation is not fixed ``canonicaly". That is, we need to
know the measure in order to represent the momentum observable.
This is in sharp contrast with the strategy followed in the
algebraic method \cite{tate}, where one first represents the
operators and later looks for a measure that renders the operators
Hermitian. It seems that, even for the simplest field theory
system, one needs to modify the strategy slightly.

Observe that we have already encountered two new actors in the play.
First comes the {\it quantum configuration space} $\overline{\cal C}$,
and the second one is the measure $\mu$ thereon. Thus, one will need to
specify these objects in the construction of the theory. To do this, we
will carry out  a two step process. First we need to find the measure
$d\mu$ on the quantum configuration space in order to get the Hilbert
space ${\cal H}_{\rm s}$ and second we need to find the multiplicative
term of the basic operator (\ref{momop}).

The strategy that seems natural to determine the measure and the
multiplicative term is to suppose that we possess a Fock
representation (it does not matter which particular one, since 
the results will be general enough). 
This representation must have a unitarily equivalent counterpart 
in the Schr\"odinger picture, and therefore fixes the
measure and the multiplicative operator in the functional
framework. That is, given a Fock representation, we want to find
the Schr\"odinger representation that is equivalent to that one.
In the remainder of this section we will dedicate
Sec.\ref{sec:3.a} to formulate this equivalence in a precise way and
Sec.\ref{sec:3.b} to complete the Schr\"odinger representation.

\subsection{Quantum algebra and states}
\label{sec:3.a}

We shall start by assuming the existence of a consistent Fock
representation of the CCR. The question we want to address now is
how to formulate equivalence between the two different
representations for the theory. The most natural way to define
this notion is through the algebraic formulation of QFT (see
\cite{haag}, \cite{baez} and \cite{wald2} for introductions). The
main idea is to formulate the quantum theory in such a way that
the observables become the relevant objects and the quantum states
are ``secondary". Now, the states are taken to ``act" on operators
to produce numbers. For concreteness, let us recall the basic
constructions needed.

The main ingredients in the algebraic formulation are two, namely:
(1) a $C^*$-algebra ${\cal A}$ of observables, and (2) states
$\omega:{\cal A}\rightarrow \C$, which are positive linear
functionals ($\omega(A^*A)\geq0\,\forall A\in{\cal A}$) such that
$\omega(1)=1$. The value of the state $\omega$ acting on the
observable $A$ can be interpreted as the expectation value of the
operator $A$ on the state $\omega$, i.e. $\la A\ra=\omega(A)$.

For the case of a linear theory, the algebra one considers is the
so-called {\it Weyl algebra}. Each generator $W(\lambda)$ of the
Weyl algebra is the ``exponentiated" version of the linear
observables (\ref{observ}), labeled by a phase space vector
$\lambda^\alpha$. These generators satisfy the Weyl relations:
\be W(\lambda)^{*}=W(-\lambda)\: , \:\:\:\:
W(\lambda_{1})W(\lambda_{2})=e^{{{i}\over{2}}
\Omega(\lambda_{1},\lambda_{2})}
W(\lambda_{1}+ \lambda_{2}) \, . \ee

The CCR $[\hat{\w}(\lambda , \cdot),\hat{\w}(\nu , \cdot)]=-i\hbar
\w(\lambda,\nu)\hat{\rm{I}}$ get now replaced by the quantum Weyl
relations where now the operators $\hat{W}(\lambda)$ belong to the
(abstract) algebra ${\cal A}$. Quantization in the old sense means
a representation of the Weyl relations on a Hilbert space.
The relation between these concepts and the algebraic construct
is given through the GNS construction that can be stated as the
following theorem \cite{wald2}:

{\it{Let ${\cal A}$ be a $C^*$-algebra with unit and let $\omega:{\cal
A}\rightarrow \C$  be a state. Then there exist a Hilbert space
${\cal H}$, a representation $\pi:{\cal A}\rightarrow L({\cal H})$
and a vector $|\Psi_0\ra\in {\cal H}$ such that, \be \omega(A)=\la
\Psi_0,\pi(A)\Psi_0\ra_{\cal H} \label{teo-expect} \, . \ee
Furthermore, the vector
$|\Psi_0\ra$ is cyclic. The triplet $({\cal H},\pi,|\Psi_0\ra)$
with these properties is unique (up to unitary equivalence)}}.

One key aspect of this theorem is that one may have different, but
unitarily equivalent, representations of the Weyl algebra, which
will yield {\it equivalent} quantum theories. This is the precise
sense in which the Fock and Schr\"odinger representations are
related to each other. Let us be more specific. We know exactly
how to construct a Fock representation from the symplectic vector
space $(V,\w_{V})$ endowed with a complex structure $J$
\cite{wald2}. The infinite dimensional freedom in choice of
representation of the CCR relies in the choice of admissible $J$,
which gives rise to the one-particle Hilbert space{\footnote{It is
worth pointing out that from the infinite possible $J$ there are
physically inequivalent representations \cite{wald2}, a clear
indication that the Stone-von
 Neumann theorem does not generalize to field theories.}}
 ${\cal{H}}_{0}$. Thereafter, the construction is completely natural
 and there are no further choices to be made: We take the Hilbert
 space of the QFT to be ${\cal{F}}_{s}({\cal{H}}_{0})$.
 The fundamental observables $\hat{\w}(\phi,\cdot)$ on
 ${\cal{F}}_{s}({\cal{H}}_{0})$ then are defined by
 $\hat{\w}(\phi,\cdot)=iA(\overline{K\phi})-iC(K\phi)$,
 where $C$ and $A$ are respectively the creation and
 annihilation operators, and $K$ is the restriction to $V$ of the
 orthogonal projection map  $\tilde{K}:V_{\mu}^{{\bf{C}}}\to
 {\cal{H}}_{0}$ in the inner product $\mu(\overline{\phi_{1}},\phi_{2})
 =-i\w(\overline{\phi_{1}},\phi_{2})$. Hence, if we suppose that we have
  a complex structure on $V$ (i.e., a Fock representation) we can now
  compute the expectation value of the Weyl operators on the Fock
  vacuum and thus obtain a positive linear functional $\omega_{\rm fock}$
  on the algebra ${\cal A}$. Now, the Schr\"odinger representation that
  will be equivalent to the Fock construction will be the one that the
  GNS construction provides for the {\it same} algebraic state
  $\omega_{\rm fock}$. Our job now is to complete the Schr\"odinger
  construction such that the expectation value of the corresponding
  Weyl operators coincide with those of the Fock representation.

The first step in this construction consists in writing the expectation
value of the Weyl operators in the Fock representation in terms of the
complex structure $J$. By hypothesis, we have a triplet
$({\cal{F}}_{s}({\cal{H}}_{0}),R_{\rm{fock}},
\w_{{\cal{F}}_{s}({\cal{H}}_{0})})$, where (i) ${\cal{F}}_{s}({\cal{H}}_{0})$
is the symmetric Fock space specified by some complex structure
(ii) $R_{\rm{fock}}$ is a map from the Weyl algebra to the
 collection of all bounded linear maps on
 ${\cal{F}}_{s}({\cal{H}}_{0})$ ($R_{\rm{fock}}$
 sends the Weyl generator $\hat{W}(\phi)$, labeled by $\phi$,
 to the operator $\exp \bigl[i\hat{\w}(\phi,\, \cdot \,)\bigr]
 \in L({\cal{F}}_{s}({\cal{H}}_{0}))$, and is extendable to the
 whole algebra by linearity and continuity) and (iii)
 $\w_{{\cal{F}}_{s}({\cal{H}}_{0})}$ is the vacuum state of the theory.
 Thus, by virtue of the GNS construction, the value of the state
 $\omega_{\rm{fock}}$ acting on the Weyl generators $\hat{W}(\phi)$
 is interpreted as the expectation value of the corresponding
 operators $R_{\rm{fock}}(\hat{W}(\phi))$ on the vacuum state
 $\w_{{\cal{F}}}$ (from now on we replace ${\cal{F}}_{s}({\cal{H}}_{0})$
 by ${\cal{F}}$):
\be
\label{expec-W}
\omega_{\rm{fock}}(\hat{W}(\phi))=\la \w_{{\cal{F}}},R_{\rm{fock}}
(\hat{W}(\phi))\w_{{\cal{F}}}\ra_{{\cal{F}}} \, . 
\ee
Now, since $R_{\rm{fock}}(\hat{W}(\phi))=\exp \bigl[i\hat{\w}(\phi,\,
 \cdot \,)\bigr]= \exp\bigl(C(K\phi)-A(\overline{K\phi})\bigr)$, we
 can rewrite, by using the Baker-Campbell-Hausdorff relation, the
 corresponding operator to the Weyl generator as follows
\be
R_{\rm{fock}}(\hat{W}(\phi))=\exp \bigl(C(K\phi)\bigr)\exp
\bigl(-A(\overline{K\phi})\bigr)\exp \bigl(-{{1}\over{2}}
[A(\overline{K\phi}),C(K\phi)]\bigr) \, . 
\ee
But the commutator $[A(\overline{K\phi}),C(K\phi)]$ is equal to
$(\overline{K\phi})_{A}(K\phi)^{A}{\hat{\rm{I}}}$. Thus, since
\be
(\overline{K\phi_{1}})_{A}(K\phi_{2})^{A}=
(K\phi_{1},K\phi_{2})_{{\cal{H}}_{0}}={{1}\over{2}}\mu(\phi_{1},\phi_{2})
-{{i}\over{2}}\w(\phi_{1},\phi_{2}) \, ,
\ee
then $(\overline{K\phi})_{A}(K\phi)^{A}={{1}\over{2}}\mu(\phi,\phi)$.
Therefore, the vacuum expectation value of $R_{\rm{fock}}(\hat{W}(\phi))$
is given by
\be
\la R_{\rm{fock}}(\hat{W}(\phi))\ra_{\rm{vac}}=\la\w_{{\cal{F}}},
\exp \bigl(C(K\phi)\bigr)\exp \bigl(-A(\overline{K\phi})\bigr)
\w_{{\cal{F}}}\ra_{{\cal{F}}}\: \exp \bigl( -{{1}\over{4}}\mu(\phi,\phi)\bigr) \, .  
\ee

Because $\la\w_{{\cal{F}}},\Psi\ra_{{\cal{F}}}=0$ for all $\Psi \in
{\cal{F}}$ such that $\Psi=(0,\phi^{A_{1}},\phi^{(A_{1}A_{2})},...,
\phi^{(A_{1}...A_{n})},...)$, then $\la\w_{{\cal{F}}},\exp
\bigl(C(K\phi)\bigr)\exp \bigl(-A(\overline{K\phi})\bigr)
\w_{{\cal{F}}}\ra_{{\cal{F}}}=\la\w_{{\cal{F}}},\w_{{\cal{F}}}
\ra_{{\cal{F}}}$. Hence, if the vacuum state is normalized,
substituting $\la R_{\rm{fock}}(\hat{W}(\phi))\ra_{\rm{vac}}$
in (\ref{expec-W}) we obtain that the value of the state
$\omega_{\rm{fock}}$ acting on the Weyl generators
$\hat{W}(\lambda)$ is given by the following expression
\be
\omega_{\rm fock}(\hat{W}(\lambda))=e^{-\frac{1}{4}\mu(\lambda,\lambda)} \, , 
\label{magic}
\ee
where, thanks to the symplectic map ${\cal{I}}_{t}$, we were able to put
$\lambda$ as a label vector for both covariant and canonical approaches.
Note that the GNS construction is precisely the technology that allows
us to invert the process. That is, from the point of view of the
algebraic approach, the choice of a complex structure $J$ defines the
Fock representation via the GNS construction based upon a state
$\omega_{\rm{fock}}$, which is defined on the basic generators of
the Weyl algebra by Eq(\ref{magic}).

\subsection{Functional representation}
\label{sec:3.b}

The next step is to complete the Schr\"odinger representation.
That is, find the measure $d\mu$ and the multiplicative term in
(\ref{momop}), that corresponds to the given Fock representation.

In order to specify the measure $\d\mu$ that defines the Hilbert
space, it suffices to consider configuration observables. Now, we
 know how to represent these observables {\it independently} of
 the measure since they are represented as multiplication operators
 as given by (\ref{comfop}). The Weyl observable $\hat{W}(\lambda)$
 corresponding to $(0,f)^\alpha$ in the Schr\"odinger picture has
 the form
\be
R_{\rm{sch}}(\hat{W}(\lambda))=e^{i\hat{\varphi}[f]} \, . 
\ee
Now, the equation (\ref{magic}) tells us that the state
$\omega_{\rm sch}$ should be such that, \be \omega_{\rm
sch}(\hat{W}(\lambda))=\exp\left[-\frac{1}{4}\mu(\lambda,\lambda)
\right]=\exp\left[-\frac{1}{4}\int_\Sigma\d^3x\,f\,B\,f\right] \, , 
\label{VEV1} \ee where we have used (\ref{pint-operadores}) in the
last step. On the other hand, the left hand side of (\ref{magic})
is the vacuum expectation value of the $\hat{W}(\lambda)$
operator. That is,
\be \omega_{\rm{sch}}(\hat{W}(\lambda))=\int_{\overline{\cal C}}\d\mu
 \, \overline{\Psi_0}(R_{\rm{sch}}(\hat{W}(\lambda))\cdot\Psi_0)=
 \int_{\overline{\cal
C}}\d\mu\,e^{i\int_\Sigma \d^3x\,f\,\varphi} \, . \label{VEV2} \ee
Let us now compare (\ref{VEV1})
and (\ref{VEV2}), \be \int_{\overline{\cal
C}}\d\mu\,e^{i\int_\Sigma \d^3x\,f\,\varphi}=
\exp\left[-\frac{1}{4}\int_\Sigma\d^3x\,f\,B\,f\right] \, . 
\label{gauss2}\ee

At this point, we take a brief {\it detour} in order to understand
the meaning of (\ref{gauss2}). Since in the case of infinite
dimensional vector spaces $\cal{V}$, the {\it Fourier Transform}
of the measure $\tilde{\mu}$ is defined as
\[
\chi_{\tilde{\mu}}(f):=\int_{\cal{V}}\d \tilde{\mu}\,e^{if(\varphi)} \, , 
\]
where $f(\varphi)$ is an arbitrary continuous function(al) on ${\cal{V}}$, it
turns out that under certain technical conditions, the Fourier
transform $\chi$ characterizes completely the measure $\tilde{\mu}$. This
fact is particularly useful for us since it allows to give a
precise definition of a Gaussian measure. Let us assume that ${\cal{V}}$
is a Hilbert space and $O$ a positive-definite, self-adjoint
operator on ${\cal{V}}$. Then a measure $\tilde{\mu}$ is said to be Gaussian if
its Fourier transform has the form,
\be
\chi_{\tilde{\mu}}(f)=\exp\left(-\frac{1}{2}\la f,Of\ra_{\cal{V}}\right)\,
,\label{gaussian}
\ee
where $\la\cdot,\cdot\ra_{\cal{V}}$ is the Hermitian inner product on
${\cal{V}}$. We can, of course, ask what the measure $\tilde{\mu}$
looks like. The
answer is that, schematically it has the form,
\be ``\d \tilde{\mu}=
\exp\left(-\frac{1}{2}\la\varphi,O^{-1}\varphi\ra_{\cal{V}}\right){\cal
D}\varphi"\, ,\label{false}
\ee
where ${\cal D}\varphi$ represents the fictitious ``Lebesgue-like"
measure on ${\cal{V}}$. The expression (\ref{false}) should be taken
with a grain of salt since it is not  completely well defined (whereas
(\ref{gaussian}) is). It is
nevertheless useful for understanding where the denomination of
Gaussian comes from. The term
$-\frac{1}{2}\la\varphi,O^{-1}\varphi\ra_{\cal{V}}$ is
(finite and) negative definite, and gives to $\tilde{\mu}$
its Gaussian character.

Thus, returning to our particular case, we note from
Eqs. (\ref{gaussian}) and (\ref{false}) that (\ref{gauss2})
tells us that the measure $\d\mu$ is Gaussian and that it
corresponds heuristically to a measure of the form,
\be
``\d\mu=e^{-\int_{\Sigma}\varphi
B^{-1}\varphi}\;{\cal D}\varphi" \, . \label{medida}
\ee

This is the desired measure. However, we still need to find the
``multiplicative term" in the representation of the momentum
operator (\ref{momop}). For that, we will need the full Weyl
algebra and Eq.(\ref{magic}). Let us denote by $K$ the Hilbert space
obtained by completing ${\cal C}$ with respect to the fiducial inner
product $(g,f):=\int_\Sigma gf$ \cite{6}. We have to compute
$\la R_{\rm{sch}}(\hat{W}(g,f))\ra_{\rm{vac}}=
\la\Psi_{0},\exp(i\hat{\varphi}[f]-i\hat{\pi}[g]) \Psi_{0}\ra$, so let
us note that we need to use the Baker-Campbell-Hausdorff relation
to separate the operators; i.e.,
\begin{equation}
\label{b-c-h-1} \exp \bigl(
i\hat{\varphi}[f]-i\hat{\pi}[g]\bigr)=\exp \bigl(
i\hat{\varphi}[f]\bigr)\exp \bigl(-i\hat{\pi}[g]\bigr) \exp
\bigl(-{{1}\over{2}} \bigl[ i\hat{\varphi}[f],-i\hat{\pi}[g] \bigr]
\bigr) \, .
\end{equation}
Given that $\bigl[\hat{\varphi}[f],\hat{\pi}[g] \bigr] = i
\int_{\Sigma} fg\, {\rm \hat{I}}$ ($\hbar =1$), then substituting
(\ref{b-c-h-1}) in (\ref{teo-expect}) and using (\ref{magic}) we have that
\begin{equation}
\label{prelim1} e^{-{{1}\over{4}}\mu_{\Gamma}((g,f),(g,f))}=\exp
\biggl( -{{i}\over{2}}\int_{\Sigma} fg
\biggr)\la\Psi_{0},\exp(i\hat{\varphi}[f])\exp(-i\hat{\pi}[g])
\Psi_{0}\ra \, , 
\end{equation}
since $\exp \bigl( -i/2 \int_{\Sigma} fg
\, {\rm \hat{I}}\bigr)\Psi_{0}=\exp \bigl( -i/2 \int_{\Sigma} fg
\bigr)\Psi_{0}$ and $\exp \bigl( -i/2 \int_{\Sigma} fg \bigr)$
does not depend on $\varphi$.

Now,
\begin{equation}
\label{desarrollo-mom-op2} \exp(-i\hat{\pi}[g])
\Psi_{0}=\exp(-i\hat{M}+\hat{d}) \Psi_{0} \, , 
\end{equation}
with
\begin{equation}
\label{opera2} -i\hat{M}\cdot \Psi = \bigg(\int_{\Sigma}\varphi
\hat{m} g\bigg) \Psi = -iM\Psi\:\:\mbox{and} \:\:\:\hat{d} \cdot
\Psi = -\int_{\Sigma} g {{\delta \Psi}\over{\delta \varphi}} \, . 
\end{equation}
Given that $[-i\hat{M},\hat{d}]\cdot \Psi=-i\bigl[\int_{\Sigma} g
 {{\delta M}\over{\delta \varphi}} \bigr]\Psi$, then using the
Baker-Campbell-Hausdorff relation, we can write the RHS of
(\ref{desarrollo-mom-op2}) as
$\exp\bigl({{i}\over{2}}\int_{\Sigma} g  {{\delta M}\over{\delta
\varphi}} \bigr)\exp(-i\hat{M})\exp(\hat{d})\Psi_{0}$, since $M$
is linear in $\varphi$, and therefore ${{\delta M}\over{\delta
\varphi}}$ does not depend on $\varphi$. On the other hand,
$\exp(\hat{d})\Psi_{0}=\Psi_{0}$ (since $\Psi_{0}$ is constant)
and $\exp(-i\hat{M})\Psi_{0}=\exp(-iM)\Psi_{0}$. Thus,
(\ref{desarrollo-mom-op2}) is
\begin{equation}
\exp(-i\hat{\pi}[g])
\Psi_{0}=\exp\biggl({{i}\over{2}}\int_{\Sigma} g  {{\delta
M}\over{\delta \varphi}} \biggr)\exp(-iM)\Psi_{0} \, . 
\end{equation}
Substituting this last expression in (\ref{prelim1}) we have that
\begin{equation}
\label{prelim2b}
e^{-{{1}\over{4}}\mu_{\Gamma}((g,f),(g,f))}=e^{-{{i}\over{2}}\int_{\Sigma}
fg }e^{{{i}\over{2}}\int_{\Sigma} g  {{\delta M}\over{\delta
\varphi}}} \int_{{\bar{\c}}}d \mu \:
e^{i\int_{\Sigma}f\varphi}e^{-iM} \, . 
\end{equation}
Using (\ref{pint-operadores}) and (\ref{opera2}) we have that
(\ref{prelim2b}) is
\begin{equation}
\label{relacion-prim2} e^{-{{1}\over{4}}\int_{\Sigma} (f B f + f A
g - g D g-g Cf )}=e^{-{{i}\over{2}}\int_{\Sigma} fg}
e^{-{{1}\over{2}}\int_{\Sigma} g \hat{m} g}\int_{{\bar{\c}}}d \mu
\: e^{i\int_{\Sigma}(f-i\hat{m}g)\varphi} \, . 
\end{equation}
From the last relation in  (\ref{relaciones2}), and using the fact
that the integral on $\bar{\c}$ is the Fourier transform with
$f\mapsto (f-i\hat{m}g) $ of the measure (\ref{medida}), we get
\begin{equation}
\label{relacion-seg2} e^{-{{1}\over{4}}\int_{\Sigma} (f B f - g D
g+2f Ag )}=e^{-{{i}\over{2}}\int_{\Sigma} fg}
e^{-{{1}\over{2}}\int_{\Sigma} g \hat{m}
g}e^{-{{1}\over{4}}\int_{\Sigma}(f-i\hat{m}g)B(f-i\hat{m}g)} \, . 
\end{equation}
That is,
\begin{equation}
\label{relacion-ter2} e^{-{{1}\over{4}}\int_{\Sigma} (f B f - g D
g+2f Ag )}=e^{-{{i}\over{2}}\int_{\Sigma} fg}
e^{-{{1}\over{2}}\int_{\Sigma} g \hat{m}
g}e^{-{{1}\over{4}}\int_{\Sigma} f B
f}e^{{{1}\over{4}}\int_{\Sigma}
(\hat{m}g)(B\hat{m}g)}e^{{{i}\over{2}}\int_{\Sigma}
(\hat{m}g)(Bf)} \, , 
\end{equation}
where we have used the first relation in (\ref{relaciones2}) to
obtain the last term. Since (\ref{relacion-ter2}) has to be valid
for all $g$ and $f$ in $K$,
then we have that
\begin{equation}
\label{gesyefes1} -\int_{\Sigma} fAg= i \int_{\Sigma}
(\hat{m}g)(Bf)-i \int_{\Sigma} fg \, , 
\end{equation}
and
\begin{equation}
\label{ges} \int_{\Sigma} gDg=\int_{\Sigma}(\hat{m}g)(B\hat{m}g)-2
\int_{\Sigma} g \hat{m} g \, . 
\end{equation}
Using the first relation in (\ref{relaciones2}), the equation
(\ref{gesyefes1}) can be rewritten as
\begin{equation}
\label{gesyefes2} \int_{\Sigma}f(A+iB \hat{m}-i{\bf{1}})g=0 \, . 
\end{equation}
In order to find $\hat{m}$ we will assume that $iB
\hat{m}-i{\bf{1}}$ is a linear operator. Given that $A$ is linear,
then $L:=A+iB \hat{m}-i{\bf{1}}$ is also linear. The equation
(\ref{gesyefes2}) should be valid for all $f$ and $g$ in $K$, then
$Lg=0$ for all $g$ in $K$ (i.e., the kernel of the operator $L$ is
all of $K$), therefore $L=0$, and
\begin{equation}
\label{eloperador} \hat{m}=B^{-1}+iB^{-1}A \, . 
\end{equation}
Note that $\hat{m}$ is (i) a linear operator from $K$ to $K \oplus
iK$ and (ii) is symmetric with respect to the inner product on
$K$, $(f,g)=\int_{\Sigma} fg$, in the sense that
 $(g,B^{-1}g')=(B^{-1} g,g')$ and
$(g,B^{-1}Ag')=(B^{-1}Ag,g')$ for all $g$ and $g'$ in $K$.

Equation (\ref{ges}) is simply a compatibility equation. If we
substitute (\ref{eloperador}) in the RHS of (\ref{ges}), we get
(using the fact that $\hat{m}$ is symmetric),
\begin{equation}
\int_{\Sigma}g(B^{-1}+iB^{-1}A)({\bf{1}}+iA)g-2\int_{\Sigma}
g(B^{-1}+iB^{-1}A)g
= - \int_{\Sigma} g(B^{-1}+B^{-1}A^{2})g = \int_{\Sigma} gDg \, , 
\end{equation}
where the last equation follows from the first relation in
(\ref{relaciones1}), which implies that $D+B^{-1}A^{2}=-B^{-1}$
and therefore $B^{-1}+B^{-1}A^{2}=-D$.

Substituting (\ref{eloperador}) in (\ref{opera2}) we get $\hat{M}$.
Thus, the representation of the
operator $\hat{\pi}[g]$, for the general case of arbitrary complex
structure (\ref{est.com}), is given by
\begin{equation}
\label{repdepi} \hat{\pi}[g]\cdot \Psi[\varphi]=-i\int_{\Sigma}
\biggl(g  {{\delta}\over{\delta \varphi}}
-\varphi(B^{-1}+iB^{-1}A)g \biggr) \Psi[\varphi] \, , 
\end{equation}
which can be rewritten in terms of the operator $C$, because
from the third relation in (\ref{relaciones1}) it follows that
$B^{-1}A=-CB^{-1}$ and consequently,
\begin{equation}
\label{repdepi-conC2} \hat{\pi}[g]\cdot
\Psi[\varphi]=-i\int_{\Sigma} \biggl(g  {{\delta}\over{\delta
\varphi}} -\varphi(B^{-1}-iCB^{-1})g \biggr) \Psi[\varphi] \, . 
\end{equation}

To summarize, we have used the vacuum expectation value condition
(\ref{magic}) in order to construct the desired Schr\"odinger
representation, namely, a unitarily equivalent representation of
the CCR on the Hilbert space defined by functionals of initial
conditions. We have provided the most general expression for the
quantum Schr\"odinger theory, for arbitrary embedding of $\Sigma$
into ${}^4M$. We saw that the only possible representation was in terms of a
probability measure, thus ruling out the naive ``homogeneous
measure". This conclusion made us realize that both the choice of
measure and the representation of the momentum operator were
intertwined; the information about the complex structure $J$ that
lead to the ``one-particle Hilbert space" had to be encoded in both
of them. We have shown that the most natural way to put this
information as conditions on the Schr\"odinger representation was
through the condition (\ref{magic}) on the vacuum expectation
values of the basic operators. This is the non-trivial input in
the construction. 

Before ending this section, several remarks are in order.
\begin{enumerate}

\item Quantum configuration space. In the introduction of
Sec.~\ref{sec:3} we made the distinction between the classical
configuration space ${\cal C}$ of initial configurations
$\varphi(x)$ of compact support and the quantum configuration
space $\overline{\cal C}$. So far we have not specified
$\overline{\cal C}$. In the case of Minkowski spacetime and flat
embeddings, where $\Sigma$ is a Euclidean space, the quantum
configuration space is the space ${\cal J}^*$ of tempered
distributions on $\Sigma$. However, in order to define this space
one uses the linear and Euclidean structure of $\Sigma$ and it is
not trivial to generalize it to general curved manifolds. These
subtleties lie outside the scope of this paper.

\item Gaussian nature of the measure.
Note that the form of the measure given by (\ref{gauss2})
is always Gaussian. This is guarantied by the fact that the
operator $B$ is positive definite in the ordinary $L^2$ norm on
$\Sigma$, whose proof is given in \cite{6}. However, the
particular realization of the operator $B$ will be different for
different embeddings $T_t$ of $\Sigma$ (cf.
Eq(\ref{part-realiza})). Thus, for a given $J$, the explicit form
of the Schr\"odinger representation depends, of course, on the
choice of embedding.

\item  Hermiticity. In order to have a consistent quantization,
one has to ensure that the operators associated to the basic
(real) observables satisfy the ``reality conditions", which in
this case means that they should be represented by Hermitean
operators. It is straightforward to show that the operator given
by (\ref{repdepi-conC2}) is indeed Hermitian.

\item Flat embedding in Minkowski spacetime. Let us now consider
the most common and simple  case, where the complex structure is
chosen to yield the standard positive-negative frequency
decomposition. This choice is associated to a constant vector
field $t^a$. Furthermore, $\Sigma$ is chosen to be the (unique)
normal to $t^a$, namely the inertial frame in which the vector
field $t^a$ is ``at rest". Thus, the complex structure $J$ is
given by
$J(\varphi,\pi)=(-(-\Delta+m^2)^{-1/2}\pi,(-\Delta+m^2)^{1/2}\varphi)$,
which means that $A=C=0$, $B=(-\Delta+m^2)^{-1/2}$ and
$D=-(-\Delta+m^2)^{1/2}$. The quantum measure is then
$``\d\mu=e^{-\int\varphi(-\Delta+m^2)^{1/2}\varphi}\;{\cal
D}\varphi"$. Thus, we recover immediately the Gaussian measure,
existing in the literature \cite{glimm,baez}, that corresponds to
the usual Fock representation. As should be clear, this represents
a very particular case (Minkowski spacetime and flat embeddings)
of the general formulae presented in this section (valid for each
globally hyperbolic spacetime and arbitrary embeddings).

\end{enumerate}

It is illuminating to further compare the resulting Schr\"odinger 
representation with its Fock couterpart. This is done in \cite{ccq2}.

\section{Discussion}
\label{sec:4}

In this paper we have constructed the Schr\"odinger representation
for a scalar field on an arbitrary, globally hyperbolic spacetime.
We have particularly emphasized the classical objects that need to
be specified in order to have these representations. It is known
that in the case of the Fock representation, formulated more
naturally in a covariant setting, the only relevant construct is
the complex structure $J$ (or alternatively, as Wald chooses to
emphasize, the metric $\mu$); the infinite freedom in the choice
of this object being precisely the ambiguity in the choice of
quantum representation for the Fock Hilbert space. In the case of
the functional representation we have, in addition to $J$, a second
classical construct, namely the choice of embedding of $\Sigma$.
Even when one has a unique well-defined theory in the Fock
language, the induced descriptions on two different embeddings
${T}_1$ and $T_2$ of $\Sigma$ might not be (unitarily) equivalent.
This second ambiguity was recently noted in \cite{mad}. This means
that there might not be a unitary operator (that is, the evolution operator
if one $\Sigma_2$ is to the future of the other surface 
$\Sigma_1$) that relates both Schr\"odinger descriptions. This
apparently general feature of QFT on curved space-times has been
recently confirmed in the quantum evolution of Gowdy $T^3$
cosmological models where the quantum description is reduced to a
scalar field on a fixed expanding background \cite{ccq,torregow}.

We have used the algebraic formulation of quantum field theory to
make precise the sense in which the Schr\"odinger representation
can be unambiguously defined. In particular, the way in which the
Fock representation is ``Gaussian" in the functional language has
been discussed in detail. We have noted that without some external
input in the construction, such as the choice of a complex
structure on the space of initial conditions, there is no a-priori
canonical way of finding a representation of the CCR; the
reality-Hermiticity conditions are not enough to select the
relevant representation and inner product. The exact implications
of this result for full canonical quantum gravity are, in our
opinion, still open. There are at least two aspects to this
question. The first one has to do with the choice of the
``physically relevant" inner product in full quantum gravity,
namely when the theory does not reduce to a model field theory.
The second aspect has to do with unitary evolution in general. In
particular, it is not clear whether a lack of unitary evolution
and existence of the Schr\"odinger representation is a serious enough obstacle to render the theory useless. This possibility has been analyzed
previously by several authors \cite{haji,ccq,torregow}. 

We hope
that the material presented here will be of some help in setting
the language for the task of understanding the fine issues of
finding the ``right" representation for, say, midisuperspace
models in quantum gravity \cite{einstein,gowdy,ccq}, and quantum 
gravity at large. 
In particular, these issues on non-unitarily
related measures have emerged in the low energy limit of semi-classical
loop quantum gravity and its relation to Fock structures \cite{loopy}.
It is important to understand these results from the broader 
perspective of curved space-time \cite{ccqn}.

\section*{Acknowledgments}

We would like to thank R. Jackiw for drawing our attention to
Refs.~\cite{reviews,jackiw} and J.M. Velhinho for correspondence.
A.C. would like to thank the hospitality
of the Perimeter Institute for Theoretical Physics, where part of this
work was completed. This work was in part supported by DGAPA-UNAM
Grant No. IN121298, by CONACyT grants J32754-E and by NSF grant
No. PHY-0010061. J.C. was supported by a UNAM (DGEP)-CONACyT
Graduate Fellowship.

\end{document}